\documentclass[10pt]{article}
\usepackage{opex2preprint}
\usepackage{cite}
\usepackage{graphicx}

\definecolor{darkgreen}{rgb}{0,0.5,0}
\definecolor{paleblue}{rgb}{0.75,0.75,1}

\begin{document}

\noindent\textbf{Preprint of:}\\
Timo A. Nieminen and Dmitri K. Gramotnev\\
``Non-steady-state extremely asymmetrical scattering of waves in periodic
gratings''\\
\textit{Optics Express} \textbf{10}, 268--273 (2002)\\
This paper is freely available at:
\url{http://www.opticsexpress.org/abstract.cfm?URI=OPEX-10-6-268}\\
which includes multimedia files not included with
this preprint version.

\hrulefill

\title{Non-steady-state extremely asymmetrical scattering of waves in periodic
gratings}

\author{Timo A. Nieminen}
\address{Centre for Biophotonics and Laser Science, Department of Physics,
The University of Queensland, Brisbane QLD 4072, Australia}
\author{Dmitri K. Gramotnev}
\address{Centre for Medical, Health and Environmental Physics, School of
Physical and Chemical Sciences, Queensland University of Technology, GPO Box
2434, Brisbane QLD 4001, Australia}

\begin{abstract}%
Extremely asymmetrical scattering (EAS) is a highly resonant type of Bragg
scattering with a strong resonant increase of the scattered wave amplitude
inside and outside the grating. EAS is realized when the scattered wave
 propagates
parallel to the grating boundaries. We present a rigorous
algorithm for the analysis of non-steady-state EAS, and investigate
the relaxation of the incident and scattered wave amplitudes to their
steady-state values. Non-steady-state EAS of bulk TE electromagnetic
waves is analyzed in narrow and wide, slanted, holographic gratings. Typical
relaxation times are determined and compared with previous rough estimations.
Physical explanation of the predicted effects is presented.
\end{abstract}

\noindent

\section{Introduction}

Extremely asymmetrical scattering (EAS) is a radically new type of Bragg
scattering in slanted, periodic, strip-like wide gratings, when the first
diffracted order satisfying the Bragg condition (scattered wave) propagates
parallel to the front grating
boundary~\cite{kishino,bakhturin,pla,jpd,deas,grazing,rigorous,quebec,vmp}.
The main characteristic feature of EAS is the strong resonant increase of
the scattered wave amplitude, compared to the amplitude of the incident wave at
the front
boundary~\cite{kishino,bakhturin,pla,jpd,deas,grazing,rigorous,quebec,vmp}.
Other unique features of EAS include additional resonances in non-uniform
 gratings~\cite{deas},
detuned gratings~\cite{quebec}, in wide gratings when the scattered wave
 propagates at a grazing angle with respect to the boundaries~\cite{grazing},
and  the unusually strong
sensitivity of EAS to small variations of mean structural parameters at the
 grating
boundaries~\cite{vmp}. The additional resonances may result in amplitudes of the
 scattered and
incident waves in the grating that can be dozens or even hundreds of times
 larger than that of the
incident wave at the front boundary~\cite{deas,grazing,rigorous,quebec,vmp}.

One of the main physical reasons for all these unusual features of EAS is the
diffractional divergence of the scattered wave inside  and outside the
grating~\cite{bakhturin,pla,jpd,deas,grazing,vmp}. Indeed, the scattered wave
results from scattering of the incident wave inside the grating, and propagates
parallel to the grating. Thus, it is equivalent to a beam located
within the grating, of an aperture equal to the grating
width. Such a beam will diverge outside the grating due to
diffraction. Therefore, steady-state EAS is formed by the two physical effects
-- scattering and diffractional
divergence. Based on the equilibrium between these processes, an approximate
analytical method of analysis of EAS, applicable to all types of waves,
has been developed~\cite{bakhturin,pla,jpd,grazing,quebec,vmp}.

A reasonable question is which of the EAS resonances are practically
achievable? Any resonance is characterized by a time of relaxation, and if this
time is too large, the corresponding resonance cannot be achieved in practice.
In the case of EAS, large relaxation times may result in excessively large
apertures of the incident beam being required for the steady-state regime
to be realized~\cite{jpd}. It is obvious that knowledge of relaxation times
and relaxation (non-steady-state) processes during EAS is essential for the
successful development of practical applications of this highly unusual type of
scattering.

Until recently, only estimates of relaxation times for EAS in uniform gratings
have been made~\cite{bakhturin,pla,jpd}. The
analysis was based on physical assumptions and speculations~\cite{pla,jpd},
rather than direct treatment of non-steady-state regime of EAS. Simple
analytical equations for the relaxation times were derived~\cite{pla,jpd}.
However, the accuracy of these equations is questionable, since their derivation
did not take into account re-scattering processes in the grating.
Moreover, these equations are not applicable in the presence of the additional
resonances~\cite{deas,grazing,quebec}.

Therefore, the first aim of this paper is to present an efficient
numerical algorithm for the rigorous analysis of non-steady-state EAS of bulk
electromagnetic waves in wide uniform and non-uniform holographic gratings. The
second aim is to investigate non-steady-state EAS and accurately determine
relaxation times in narrow and wide uniform periodic gratings. In particular,
amplitudes of the incident wave (0th diffracted order) and scattered wave
(+1~diffracted order) inside and outside the grating will be
determined as functions of time and coordinates after the incident wave is
``switched on.''

\section{Structure and methods of analysis}

\begin{figure}[htb]
\begin{center}
\includegraphics[width=0.47\textwidth]{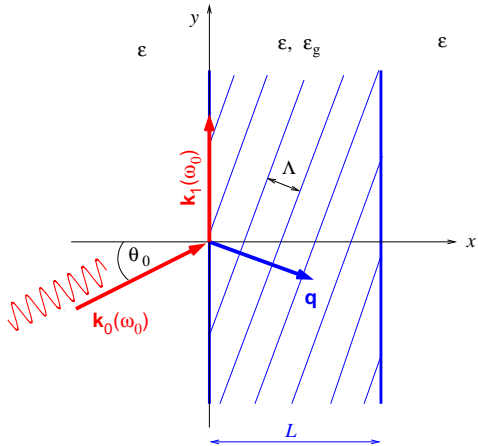}
\end{center}
\vspace{-5mm}
\caption{The geometry of EAS in a slanted periodic grating.}
\label{gratingfig}
\end{figure}

Consider an isotropic medium with a slab-like, uniform, holographic grating
with a sinusoidal modulation of the dielectric permittivity (see
fig~\ref{gratingfig}):
\begin{equation}
\begin{array}{rcll}
\epsilon_\mathrm{s}  & = & \epsilon +
\epsilon_\mathrm{g} \exp( \mathrm{i} q_x x + \mathrm{i} q_y y ) +
\epsilon_\mathrm{g}^\star \exp( - \mathrm{i} q_x x - \mathrm{i} q_y y ),
& \textrm{if}\; 0 < x < L,\\
\epsilon_\mathrm{s} & = & \epsilon,& \textrm{if}\; x < 0 \;\textrm{or}\; x > L,
\end{array}
\label{gratingeqn}
\end{equation}
where $L$ is the grating width, $\epsilon_\mathrm{g}$ is the grating amplitude,
the mean dielectric permittivity $\epsilon$ is the same inside and outside the
grating, $q_x$ and $q_y$ are the $x$ and $y$ components of the reciprocal
lattice vector $\mathbf{q}$, $q = 2\pi/\Lambda$, $\Lambda$ is the period of the
grating, and the coordinate system is shown in figure~\ref{gratingfig}. There
is no dissipation in the medium ($\epsilon$ is real and positive), and the
structure is infinite along the $y$ and $z$ axes.

Non-steady-state EAS in this structure occurs when the incident wave is
switched on at some moment of time. Then, both the incident
and scattered wave amplitudes inside and outside the grating evolve in time and
gradually relax to their steady-state values at $t = + \infty$. This occurs when
 the incident pulse, having
an infinite aperture along the $y$ and $z$ axes, is the product of a sinusoid
 and a step function of time.

However, the numerical analysis of this infinitely long (in time) pulse is
 inconvenient,
since its Fourier transform contains a $\delta$-function. Therefore, in order
to calculate non-steady-state amplitudes of the incident and scattered waves in
the structure at an arbitrary moment of time $t = t_0$, we consider a
rectangular (in time) sinusoidal incident pulse of time length $2t_0$,
amplitude  $E_{00}$, and angle of
incidence $\theta_0$ (fig.~1), which has a simple analytical Fourier transform.
We also assume that the incident pulse is switched on instantaneously (with the
amplitude $E_{00}$) everywhere in the grating. That is, we
ignore the process of propagation of the pulse through the grating.
This assumption allows a substantial reduction in computational time. It
is correct only if the time $t_0$, at which non-steady-state
amplitudes are calculated, is significantly larger than $t_\mathrm{g} =
(L/\cos\theta_0)/(c\epsilon^{-1/2})$ -- the time of propagation of the incident
pulse across the grating. We will see below that this condition is satisfied
(for most times) since typical relaxation times $\tau \gg t_\mathrm{g}$.

Thus, at $t = 0$, the front of the pulse is at the rear grating boundary,
i.e. the incident field is zero behind the grating (at $x > L$). This means that
if the angle of incidence $\theta_0 \ne 0$ (fig.~\ref{gratingfig}), then the
amplitude of the incident pulse is not constant along a wave front. Therefore,
the incident pulse experiences diffractional divergence, and the spatial Fourier
transform should be used. However, this divergence can be ignored in our further
considerations, since it may be significant only within several wavelengths near
the front of the pulse, i.e. for times $\approx 10^{-14}$\,s or smaller
(this is the time for the pulse to travel a distance of a few wavelengths). This
time is several orders of magnitude less than typical relaxation times (see
below). Therefore, only for very short time intervals after switching an
incident pulse on can noticeable errors result from the above approximations.

The frequency spectrum of this input is determined from its analytical
Fourier transform. As a result, the incident pulse is represented by a
superposition of an infinite number of plane waves having different frequencies
and amplitudes, all incident at $\theta_0$. Note that the spectrum is the
incident pulse depends on the pulse width $2t_0$, and is different for every
time $t_0$ at which we calculate the fields.
The steady-state response of the
grating to each plane wave is determined by means of the
rigorous theory of steady-state EAS~\cite{rigorous}, based on the enhanced
T-matrix algorithm~\cite{moharam1995a,moharam1995b}, or the approximate
theory~\cite{quebec} (if applicable). Thus the
frequency spectrum of the incident and scattered waves inside and outside the
grating is obtained, and their amplitudes as functions of the
$x$-coordinate at any moment time can be found using the inverse Fourier
 transform. Due
to the geometry of the problem, the non-steady-state incident and
scattered wave amplitudes do not depend on the $y$-coordinate.

Note that the inverse Fourier transform is taken at $t = t_0$, i.e. at the
middle of the incident pulse, in order to minimize numerical errors. The
inverse Fourier transform is found by direct integration, to allow a
non-uniform set of frequency points to be used. The rapid variation of the
frequency spectra at certain points, and the wide frequency spectrum of the
input for small $t_0$ make it infeasible to use the fast Fourier transform.
The calculations are carried out
separately for each moment of time. Therefore, there is no accumulation of
numerical errors, and the errors that are noticeable at small times $\approx
10^{-14}$\,s (see above) disappear at larger times.

This approach is directly applicable for all shapes of the incident pulse, as
well as for an incident beam of finite aperture. However, for beams of finite
aperture, we should also use the spatial Fourier integral.

Since the approximate theory is directly applicable for all types of waves
(including surface and guided waves in periodic groove
arrays~\cite{bakhturin,pla,vmp}), the developed approach to
non-steady-state scattering is also applicable for all types of waves (if used
together with the approximate theory).

\section{Numerical Results}

Using the described numerical algorithm, non-steady-state EAS of bulk TE
electromagnetic waves in uniform holographic gratings given by
eqn~(\ref{gratingeqn}) has been analyzed. The grating parameters are as
follows: $\epsilon = 5$, $\epsilon_\mathrm{g} = 5 \times 10^{-3}$, $\theta_0 =
45^\circ$, and the wavelength in vacuum (corresponding to the central frequency
$\omega_0$ of the spectrum of the incident pulse) $\lambda_0 = 1\,\mu$m. The
Bragg condition is assumed to be satisfied precisely for the first diffracted
order at $\omega = \omega_0$:
\begin{equation}
\mathbf{k}_1(\omega_0) = \mathbf{k}_0(\omega_0) - \mathbf{q},
\end{equation}
where $\mathbf{k}_0(\omega)$ are the frequency dependent wave vectors of the
plane waves in the Fourier integral of the incident pulse,
$\mathbf{k}_1(\omega)$ are the wave vectors of the corresponding first
diffracted orders (scattered waves), $k_1(\omega_0) = k_0(\omega_0) =
\omega_0 \epsilon^{1/2}/c$, $\mathbf{k}_1(\omega_0)$ is parallel to the
grating boundaries (fig.~\ref{gratingfig}), and $c$ is the speed of light. Note
that if $\omega \ne \omega_0$, $\mathbf{k}_1(\omega)$ is not parallel to the
grating boundaries~\cite{quebec}.

\begin{figure}[htb]
\begin{center}
\begin{tabular}{lll}
\small\textbf{\textsf{(a)}}&\small\textbf{\textsf{(b)}}&
\small\textbf{\textsf{(c)}}\\
\includegraphics[width=0.3\textwidth]{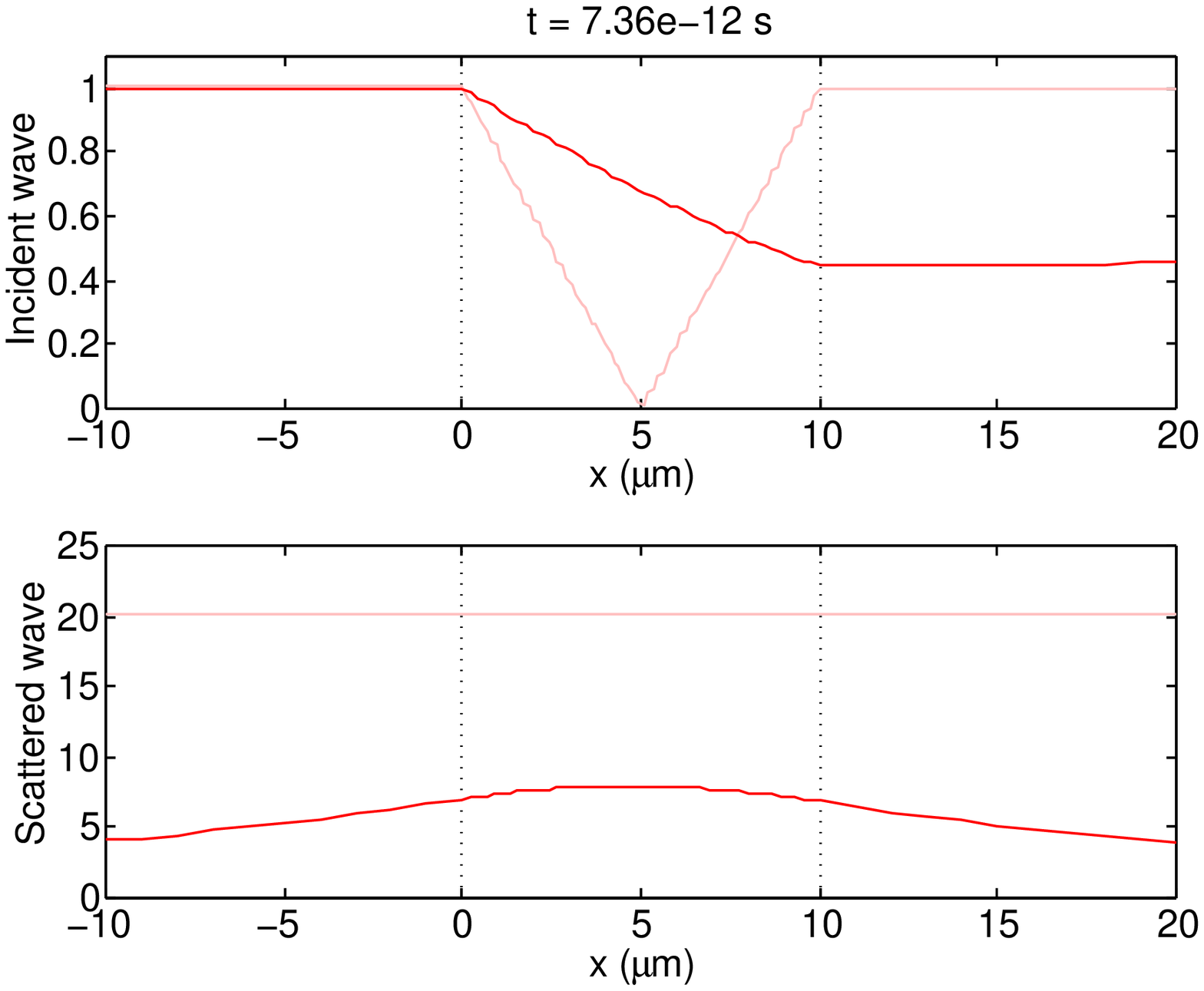}&
\includegraphics[width=0.3\textwidth]{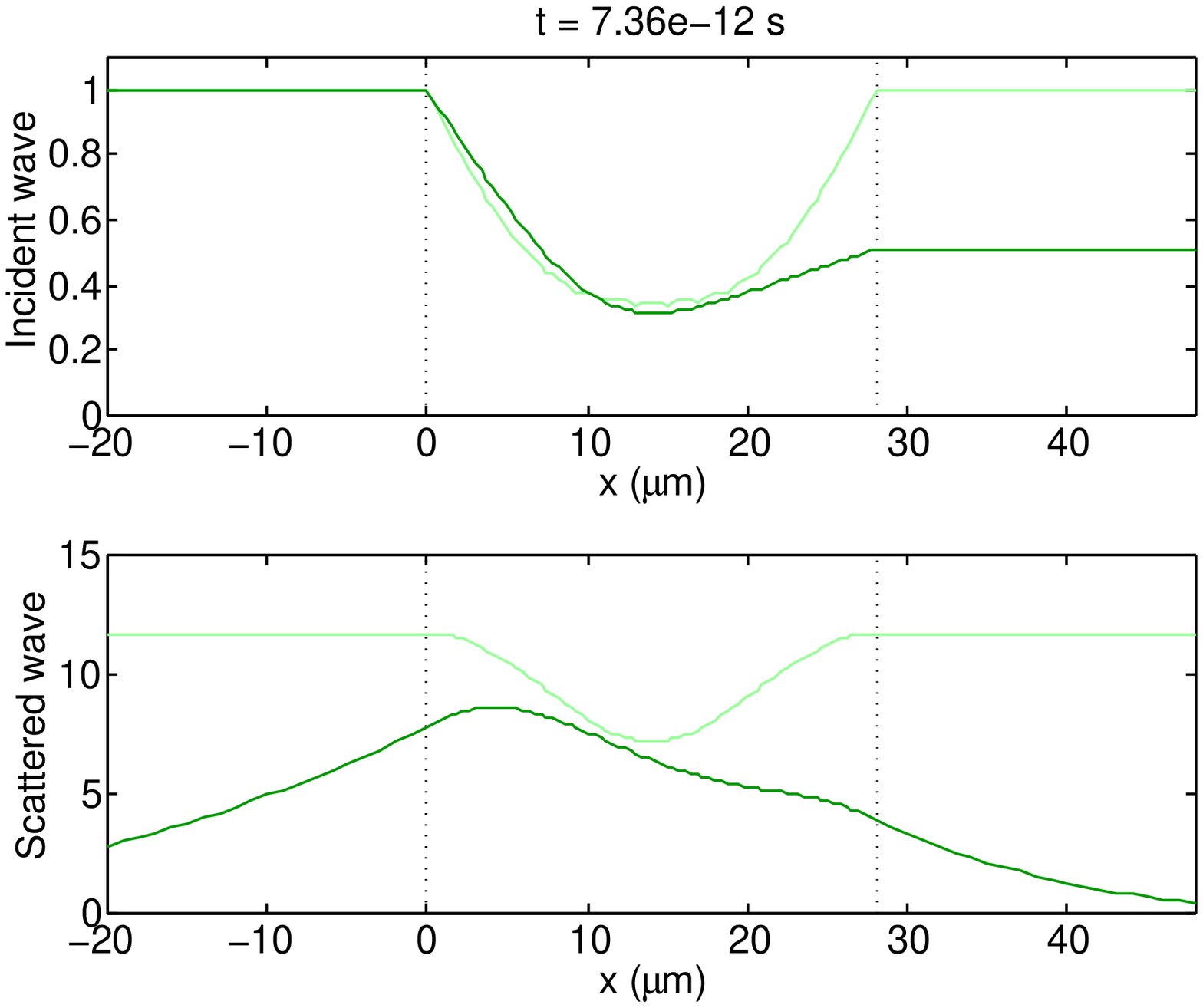}&
\includegraphics[width=0.3\textwidth]{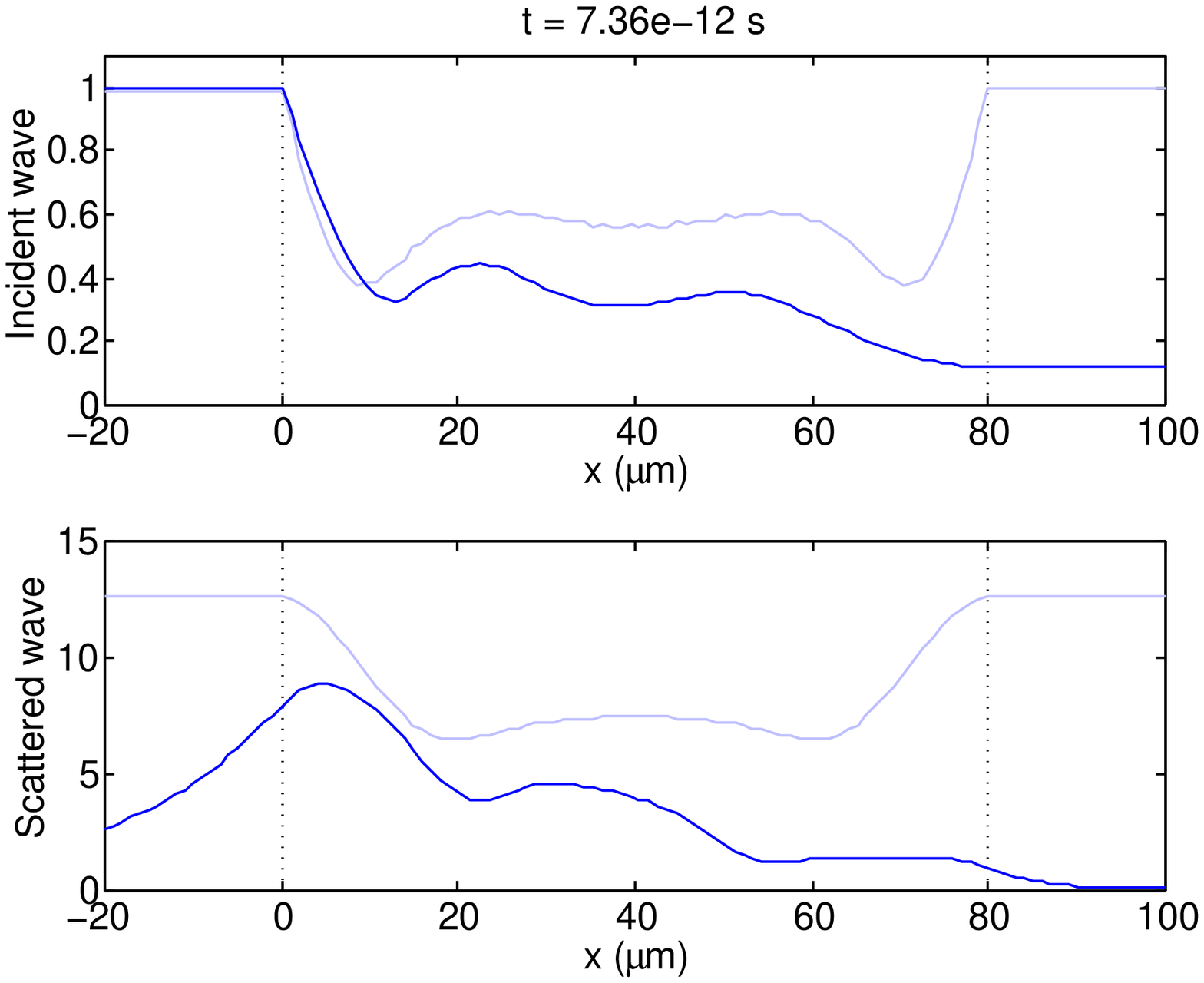}\\
\hspace*\fill {\sf\small [animated gif, 469kb]} \hspace*\fill &
\hspace*\fill {\sf\small [animated gif, 543kb]} \hspace*\fill &
\hspace*\fill {\sf\small [animated gif, 628kb]} \hspace*\fill
\end{tabular}
\end{center}
\vspace{-4mm}
\caption{Animations showing the approach of amplitudes of the scattered
(bottom graphs) and incident (top graphs) waves to the steady-state solutions
(light lines). The grating widths are \textbf{(a)} $L = 10\,\mu$m, \textbf{(b)}
$L = 28\,\mu$m $\approx L_\mathrm{c}$, and \textbf{(c)} $L = 80\,\mu$m. The
vertical dotted lines show the grating boundaries.}
\label{moviesfig}
\end{figure}

\begin{figure}[htb]
\begin{center}
\includegraphics[width=\textwidth]{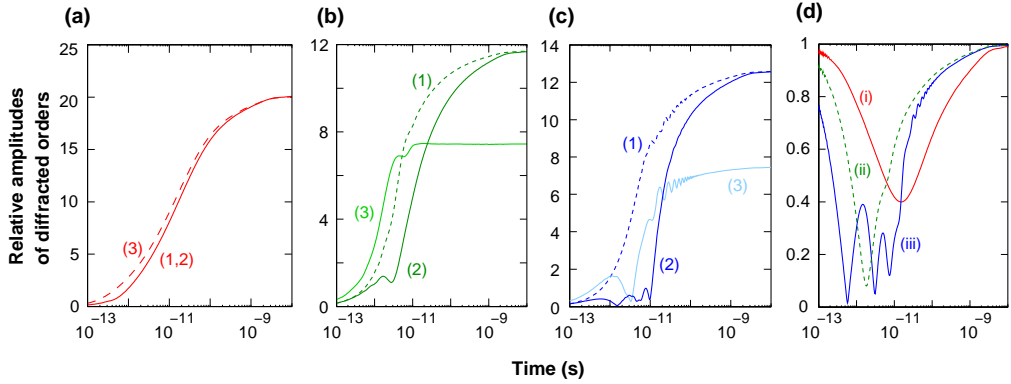}
\end{center}
\vspace{-4mm}
\caption{The time dependencies of normalized non-steady-state amplitudes of
\textbf{(a--c)} the first diffracted order (scattered wave) $|E_1/E_{00}|$, and
\textbf{(d)} the zeroth diffracted order (transmitted wave) $|E_0(x=L)/E_{00}|$.
The grating widths are $L = 10\,\mu$m (\textbf{(a)} and curve (i) in
\textbf{(d)}), $L = 28\,\mu$m $\approx L_\mathrm{c}$ (\textbf{(b)} and curve
(ii) in \textbf{(d)}), and $L = 80\,\mu$m (\textbf{(c)} and curve (iii) in
\textbf{(d)}). The scattered wave amplitudes
\textbf{(a--c)} are shown at (1) the front boundary ($x = 0$), (2) the
rear boundary ($x = L$), and (3) the middle of the grating ($x = L/2$).}
\label{timegraphs}
\end{figure}

Typical relaxation of
amplitudes of the scattered wave (+1 diffracted order) and incident wave
(0th diffracted order) over time inside and outside the gratings of
widths $L = 10\,\mu$m, $28\,\mu$m, and $80\,\mu$m is
shown in the animations in fig.~\ref{moviesfig}. The
time dependencies of non-steady-state amplitudes of the scattered wave
(+1~diffracted order) at $x = 0$, $L/2$, and $L$, and the transmitted wave (0th
diffracted order at $x = L$)  are shown in fig.~\ref{timegraphs}
for the same grating widths. Note that $L = 28\,\mu$m is
approximately equal to the critical width $L_\mathrm{c}$~\cite{deas}.
Physically, $L_\mathrm{c}/2$ is the distance within which the scattered wave can
be spread across the grating by means of the diffractional divergence, before
being re-scattered by the grating~\cite{deas}. All the curves in figures
\ref{moviesfig} and \ref{timegraphs} can equally be regarded as approximate or
rigorous, since both the theories in the considered structure give practically
indistinguishable results~\cite{rigorous}.

If $L < L_\mathrm{c}$, then the relaxation at the front and
rear grating boundaries occurs practically simultaneously (see
figures~\ref{moviesfig}(a) and \ref{timegraphs}(a)). In the middle of the
grating, the scattered wave amplitude grows slightly faster at small $t$. This
is due to energy losses from the scattered wave, caused by diffractional
divergence of the wave near the boundaries (the edges of the scattered
beam). This effect becomes more obvious with increasing grating width, and is
especially strong if $L \approx L_\mathrm{c}$ (fig.~\ref{timegraphs}(b)). This
is because the effect of the diffractional divergence in the middle of the
grating (in the middle of the beam) becomes weaker with increasing grating width
(i.e. beam aperture). At the same time, at the edges of the beam (at the
grating boundaries) the divergence is strong, resulting in a significant
reduction of the rate of change of the non-steady-state scattered wave
amplitudes (compare curves (1) and (2) with curve (3) in
fig.~\ref{timegraphs}(b)). However, in wide gratings (with $L > L_\mathrm{c}$),
the relaxation occurs first near the front boundary, and then the steady-state
scattered wave amplitudes start spreading towards the rear boundary
 (fig.~\ref{moviesfig}(c)). Therefore,
the relaxation process in the middle of the grating and especially at the rear
boundary tends to slow down compared to that near the front boundary -- compare
curves (1) and (2) in figs.~\ref{timegraphs}(b,c).

The relaxation near the front boundary takes place more or less smoothly,
except for some not very significant oscillations in wide gratings near the end
of the relaxation process (curves (1) in figs.~\ref{timegraphs}(a--c)). The same
happens in the middle of the grating and at the rear boundary in narrow gratings
($L < L_\mathrm{c}$). However, if $L \geq L_\mathrm{c}$, the relaxation curves
in the middle of the grating and at the rear boundary display a complex and
unusual behavior at small and large time intervals (see curves (2) and (3) in
figs.~\ref{timegraphs}(b,c)).

The unusual and complex character of relaxation processes in wide gratings is
 especially
obvious from the time dependencies of non-steady-state incident (transmitted)
wave amplitude at the rear grating boundary
 (figs.~\ref{moviesfig}(c),~\ref{timegraphs}(d)). In
wide gratings, these dependencies are characterized by significant oscillations
with minima that are close to zero -- see fig.~\ref{moviesfig}(c) and curve
(iii) in fig.~\ref{timegraphs}(d). The typical number of these oscillations
increases with increasing grating width (compare curves (ii) and (iii) in
fig.~\ref{timegraphs}(d)). The minima in these oscillations tend to zero with
increasing $L$. When the amplitude of the transmitted wave at the rear grating
boundary is close to zero, almost all energy of the incident wave is transferred
to the scattered wave. In wide gratings this may happen several times during the
relaxation process (figs.~\ref{moviesfig}(c),~\ref{timegraphs}(d)).

The relaxation times at the grating boundaries and at $x = L/2$, determined as
the times at which the amplitudes reach $(1 - 1/\mathrm{e})$ of their
steady-state values, are:
\begin{center}
\begin{tabular}{cccc}
$L$ & $\tau|_{x = 0}$ & $\tau|_{x = L/2}$ & $\tau|_{x = L}$ \\
%\hline
$10\,\mu$m & $4 \times 10^{-11}$\,s &
$4 \times 10^{-11}$\,s & $4 \times 10^{-11}$\,s \\
$28\,\mu$m & $6 \times 10^{-12}$\,s &
$1.7 \times 10^{-12}$\,s & $2.4 \times 10^{-11}$\,s \\
$80\,\mu$m & $7.4 \times 10^{-12}$\,s &
$1.2 \times 10^{-11}$\,s & $3.4 \times 10^{-11}$\,s
\end{tabular}
\end{center}
The relaxation times for narrow gratings, determined by means of the
developed algorithm, are about three times smaller than those
previously estimated~\cite{jpd}. The significant overestimation of
relaxation times in paper~\cite{jpd} is due to not taking into account the
effects of re-scattering of the scattered wave in the grating. Re-scattering
reduces the transmitted wave amplitude during the relaxation process
(fig.~\ref{timegraphs}(d)). Thus, the energy flow into the scattered wave is
increased, and the relaxation times are reduced (for more detailed discussion of
re-scattering see~\cite{deas}).

During the process of relaxation, the scattered wave propagates a particular
distance along the $y$-axis. Therefore, the relaxation times determine critical
apertures $l_\mathrm{c}$ (along the $y$-axis) of the incident beam, that are
required for steady-state EAS to be achieved (see also [4]). For example, for
fig.~\ref{timegraphs}(a) (with the largest values of $\tau$), the critical
aperture of the incident beam is $l_\mathrm{c} =
c\tau\epsilon^{-1/2}\cos\theta_0 \approx 0.4$\,cm, which does not present any
problem in practice.

\section{Conclusions}

This paper has developed an efficient numerical algorithm for the approximate
and rigorous numerical analysis of non-steady-state EAS of waves in uniform
slanted gratings. An unusual type of relaxation with strong oscillations of the
incident and scattered wave amplitudes has been predicted for gratings wider
than the critical width.

If used in conjunction with the approximate theory of steady-state
EAS~\cite{bakhturin,pla,jpd,quebec}, the developed algorithm is
immediately applicable for the analysis of non-steady-state EAS of all types of
waves, including surface and guided modes in periodic groove arrays.

Typical relaxation times have been calculated for narrow and wide gratings. It
has been shown that these times are significantly smaller than previous
estimates~\cite{pla,jpd}. The corresponding critical apertures of the 
incident beam that are required for achieving steady-state EAS have also been
determined. The obtained results demonstrate that steady-state EAS can readily
be achieved in practice for reasonable beam apertures and not very long
gratings.

% The commands, \begin{OEReferences} and \end{OEReferences}
% format the References section according to OpEx standard
% style, showing the title "References".
%
% The commands, \begin{OERefLinks} and \end{OERefLinks}
% format the References section according to OpEx standard
% style, if the references also include URLs or other
% unreviewed links.  In this case the title of the section
% is "References and unreviewed links".
%
%\begin{OEReferences}

%\end{OEReferences}

\end{document}